\documentclass[aps,pra,reprint,superscriptaddress,showpacs]{revtex4-1}
\usepackage{amsmath}
\usepackage{mathrsfs}
\usepackage{array}
\usepackage{graphicx}
\usepackage{amstext}
\usepackage{amsfonts}
\usepackage{bm}
\usepackage{amssymb}

\begin{document}

\title{Buildup of  Fano resonances in the time domain in a double quantum dot Aharonov-Bohm interferometer}
\author{Pei-Yun Yang}
\affiliation{Department of Physics and Center for Quantum
information Science, National Cheng Kung University, Tainan 70101,
Taiwan}
\author{Wei-Min Zhang}
\email{wzhang@mail.ncku.edu.tw}
\affiliation{Department of Physics
and Center for Quantum information Science, National Cheng Kung
University, Tainan 70101, Taiwan}

\begin{abstract}
In this paper, we investigate transient quantum transport in
a nanoscale Aharonov-Bohm (AB) interferometer consisting of a laterally coupled
double quantum dot (DQD) coupled to the source and drain electrodes.
The transient linear conductance is derived at finite temperature of the leads, and is divided into three terms contributed from different transport channels.
By observing the transient linear conductance and time evolution of the elements of the
reduced density matrix of the DQD system, we show the nature of how Fano resonance is built up in the time domain.
\end{abstract}
 \maketitle

\section{Introduction}
Fano resonance \cite{Fano1961} is a universal phenomenon throughout nuclear,
atomic, molecular and optical physics, as well as in various condensed matter systems \cite{Miroshnichenko2010}.
It is induced by interference of transport in alternative paths,
when a discrete state interacts with a continuum sharing the same energy.
Distinct from Breit-Wigner resonance \cite{Breit1936},
which arises due to counter transport in the same quantum path and displays a Lorentz line shape,
Fano resonance shows an asymmetric line profile that has both constructive
and destructive interference around the resonance energy.
In a decade ago, with the advent of femtosecond laser,
the time-dependent formation of asymmetric Fano line shapes in absorption spectra
for a photonionization process was proposed for pump-probe experiments \cite{Wickenhauser2005}.
Until very recently, such propose has been verified experimentally  in the prototype system of helium \cite{Kaldun2016}.
On the other hand, many recent experimental developments allow one to measure transient quantum transport
in different nano and quantum devices \cite{Lu2003,Bylander2005,Gustavsson2008},
thus the buildup of Fano resonance resolved in time in solid-state electronic systems is also of great interests.
In this work, we investigate the transient quantum transport though an
Aharonov--Bohm (AB) interferometer containing double quantum dots (DQDs) in parallel to explore
the buildup of Fano resonance in electronic systems.

In fact, there are more and more observations of Fano effects in novel solid state systems,
such as open quantum dots \cite{Kastner1993,Gores2000,Zacharia2001,Johnson2004},
AB interferometers \cite{Aharonov1959,Kobayashi2002,Kobayashi2003},
two-dimensional (2D) electronic waveguides and nanotubes \cite{Kim2003},
where alternative electronic paths can be achieved. Among which,
the first tunable Fano resonance is observed in an AB interferometer
containing a quantum dot in one of the two arms \cite{Kobayashi2002,Kobayashi2003},
with the benefit that the width ($\gamma$) and Fano factor ($q$) of Fano resonance are controllable by the AB phase.
Inspired by the experimental realization of DQDs embedded
into opposite arms of an AB interferometer \cite{Holleitner2001,Hatano2004,Sigrist2004},
several works have predicted Fano resonance in this system without
\cite{Kubala2002,Ladron2003,Kang2004,Bai2004,Lu2005,Kubo2006} and with \cite{Dong2004,Lopez2005} Coulomb interactions.
In particular, Kang and Cho \cite{Kang2004} revealed analytically that in the double quantum dot AB interferometer, there are two resonances in the steady-state
electron conductance at zero temperature near the bonding and antibonding states,
which are composed of a Breit-Wigner resonance and a Fano resonance. Furthermore, Kubo et al. \cite{Kubo2006}
studied the associated resonances with varying the indirect coupling between double dots through the leads,
and they found that Fano resonance can be suppressed as the indirect coupling strength decreases,
which remains two Breit-Winger peaks in the steady-state conductance.

Using the quantum transport theory based on the master equation approach \cite{Tu2008,Jin2010,Yang2017},
we are able to investigate the transient quantum transport in these novel solid-state systems,
and then can explore how Fano resonance builds up in the time domain.
We derive the transient linear conductance and the elements of reduced density matrix not only at zero
temperature but also at finite temperature.
In the situation of a symmetric setup of the double quantum dot AB interferometer, the conductance can be divided into three terms which correspond respectively
to electron transport through the bonding and the antibonding channels and the interference between these two channels.
By observing the time evolution of each term contributing to the linear conductance,
we obtain the different time scales for the formation of Fano resonance and Breit-Wigner resonance.
Moreover, we investigate the influence of the thermal effects on the time
scales to the formation of the resonances and resonance profiles.
We also compute the time evolution of the reduced density matrix elements of the DQD system.
The results help us to understand the nature of the formation of various resonances
through the transient quantum transport processes.
Also, our results reproduce the steady-state results that have been obtained
at zero temperature in the previous works \cite{Kubala2002,Ladron2003,Kang2004,Bai2004,Lu2005,Kubo2006}.
We expect that the results of the buildup of Fano resonance in the time-domain  can be observed in experiments.

The rest of paper is organized as follows.
The setup of the system, the transient transport based on master equation and the transient linear conductance
in the system of a double quantum dot AB interferometer are derived in Sec.II.
The transient dynamics of the linear conductance and the corresponding reduced density matrix of the double dots
in the symmetric situation in the molecular basis is investigated in Sec. III.
The time scale of the formation of Fano resonance is given. Also, the relation between the resonances and the reduced density matrix is discussed.
Finally, a summary is presented in Sec. IV.

\section{Double quantum dot AB interferometer}
The Hamiltonian of a nanoscale AB interferometer consisting
of two coupled single-level QDs coupled two leads is given by
\begin{align}
H=H_{DQD}+H_{E}+H_{T},
\end{align}
where $H_{DQD}$ and $H_{E}$ are respectively the Hamiltonian of the DQDs and the two leads:
\begin{align}
&H_{DQD}=\sum_{i=1}^{2}\varepsilon_{ij}d_{i}^{\dagger}d_{j}, \\
&H_{E}=\sum_{\alpha=L,R}\sum_{k}\epsilon_{\alpha k}c_{\alpha k}^{\dagger}c_{\alpha k},
\end{align}
and $d_{i}\;(d_{i}^{\dagger})$ and $c_{\alpha k}\;(c_{\alpha k}^{\dagger})$
are respectively the annihilation (creation) operators of electrons
in the $i$th dot and the $k$th level in lead $\alpha$, with
$\varepsilon_{ij}$ and $\epsilon_{\alpha k}$ being the corresponding energy.
The Hamiltonian $H_{T}$ describes the tunnelings between the dots and the leads.
\begin{align}
H_{T}=\sum_{\alpha=L,R}\sum_{i=1}^{2}\sum_{k}(V_{i\alpha k}d_{i}^{\dagger}c_{\alpha k}+h.c).
\end{align}
where $V_{i\alpha k}=\bar{V}_{i\alpha k}e^{i\phi_{i\alpha}}$ with the phase coming from
the magnetic flux $\Phi$ threading into the AB ring.
It gives $\phi_{1L}-\phi_{1R}+\phi_{2R}-\phi_{2L}=2\pi \Phi/\Phi_{0}=\varphi$,
where $\Phi_0$ is the flux quanta.
Here, we focus on the single-particle interference properties and disregard Coulomb interactions, which can be done by
controling the energy scale of the nanodevice to let the interdot Coulomb correlations
become negligible and to set up the DQDs in the Coulomb blockage regime for practical applications.

\subsection{Exact master equation approach for transient quantum transport}
The device described above can be treated as an open quantum system.
The dynamics of the DQD system is described by the reduced density matrix $\rho(t)=\mathrm{Tr}_{E}\left[ \rho_{tot}(t)\right]$, which is obtained by tracing over all
the degrees of freedom of the reservoirs from the total density matrix $\rho_{tot}(t)$ of the DQDs plus the leads.
We have derived the exact master equation which governs the dynamics of $\rho(t)$ for the DQDs \cite{Tu2008,Jin2010,Tu2012,Liu2016,Yang2017}:
\begin{align}
\frac{d\rho(t)}{dt} = & -\frac{i}{\hbar}\left[H_{S}^{\prime}(t),\rho(t)\right] +  \sum_{i,j} \Big\{\gamma_{ij}(t) \big[2d_{j}\rho(t)d_{i}^{\dagger} \nonumber \\
 & -d_{i}^{\dagger}d_{j}\rho(t)-\rho(t)d_{i}^{\dagger}d_{j}\big]  + \widetilde{\gamma}_{ij}(t)\big[d_{i}^{\dagger}\rho(t)d_{j} \nonumber \\
 & -d_{j}\rho(t)d_{i}^{\dagger}+d_{i}^{\dagger}d_{j}\rho(t)-\rho(t)d_{j}d_{i}^{\dagger}\big]\Big\}. \label{master eq}
\end{align}
The first term of Eq.~(\ref{master eq}) describes the unitary evolution of electrons in the device system,
where the renormalized Hamiltonian $H_{S}^{\prime}(t)=\sum_{i,j}\varepsilon_{ij}^{\prime}(t)d_{i}^{\dagger}d_{j}$
with the renormalized energy matrix  $\varepsilon_{ij}^{\prime}(t)$ of the DQDs
includes the energy shift of each level and the lead-induced couplings between different levels.
The remaining terms give the non-unitary dissipation and fluctuation processes
induced by back-actions of the electrons from the leads,
and are described by the dissipation and fluctuation coefficients $\boldsymbol{\gamma}(t)$ and $\widetilde{\boldsymbol{\gamma}}(t)$, respectively.
Explicitly, all the time-dependent coefficients in Eq.~(\ref{master eq}) are found to be,
\begin{subequations}
\begin{align}
& \boldsymbol{\varepsilon}^{\prime}(t) =
  \frac{1}{2}\big[\dot{\boldsymbol{u}}(t,t_0)\boldsymbol{u}^{-1}(t,t_0)- {\rm H.c}\big],
\\
& \boldsymbol{\gamma}(t)  =  -\frac{1}{2}\big[\dot{\boldsymbol{u}}(t,t_0)\boldsymbol{u}^{-1}(t,t_0)+{\rm H.c}\big], \\
& \widetilde{\boldsymbol{\gamma}}(t)  = \dot{\boldsymbol{v}}(t,t) \!-\! \big[\dot{\boldsymbol{u}}(t,t_0)\boldsymbol{u}^{-1}(t,t_0)\boldsymbol{v}(t,t)+{\rm H.c}\big],
\end{align}
\end{subequations}
where functions $\boldsymbol{u}(\tau,t_0)$ and $\boldsymbol{v}(\tau,t)$ satisfy the following integro-differential equations \cite{Tu2008,Jin2010,Yang2017},
\begin{subequations}\label{uv}
\begin{align}
&\frac{d}{d\tau}\boldsymbol{u}(\tau,t_0)+i\boldsymbol{\varepsilon}\boldsymbol{u}(\tau,t_0)+\sum_{\alpha}\int^{\tau}_{t_0}d\tau'
\boldsymbol{g}_{\alpha}(\tau,\tau')\boldsymbol{u}(\tau',t_0)=0, \\
&\frac{d}{d\tau}\boldsymbol{v}(\tau,t)+i\boldsymbol{\varepsilon}\boldsymbol{v}(\tau,t)+\sum_{\alpha}\int^{\tau}_{t_0}d\tau'
\boldsymbol{g}_{\alpha}(\tau,\tau')\boldsymbol{v}(\tau',t) \notag\\
&~~~~~~~~~~~~~=\sum_{\alpha}\int^t_{t_0}d\tau'\widetilde{\boldsymbol{g}}_{\alpha}(\tau,\tau')\boldsymbol{u}^\dag(t,\tau'),
\end{align}
\end{subequations}
subject to the boundary conditions $\boldsymbol{u}(t_0,t_0)=\boldsymbol{1}$ and $\boldsymbol{v}(t_0,t)=\boldsymbol{0}$.
In the above equations, the time non-local integral kernels, $\boldsymbol{g}_\alpha(\tau,\tau')$ and $\widetilde{\boldsymbol{g}}_{\alpha}(\tau,\tau')$
characterizing all the memory effects between the dots and lead $\alpha$ are defined as,
\begin{subequations}
\begin{align}
&\boldsymbol{g}_{\alpha}(\tau,\tau')=\int\frac{d\omega}{2\pi}\boldsymbol{\Gamma}_{\alpha}(\omega)e^{-i\omega(\tau-\tau')}, \\
&\widetilde{\boldsymbol{g}}_{\alpha}(\tau,\tau')=\int\frac{d\omega}{2\pi}\boldsymbol{\Gamma}_{\alpha}(\omega)f_{\alpha}(\omega)e^{-i\omega(\tau-\tau')},
\end{align}
\end{subequations}
where $\boldsymbol{\Gamma}_\alpha(\omega)$ is the spectral density (level broadening) of lead $\alpha$,
\begin{align}
\Gamma_{\alpha i j}(\omega)=2\pi\sum_k V_{i\alpha k}V^*_{j\alpha k}\delta(\omega-\epsilon_{\alpha k}),
\end{align}
and $f_{\alpha}(\omega)=1/(e^{\beta(\omega-\mu_\alpha)}+1)$ is the Fermi-Dirac distribution function of lead $\alpha$ at initial time $t_0$,
and $\beta=1/{k_BT}$, the inverse temperature.
Solving the inhomogeneous equation (\ref{uv}b) with the initial condition $\boldsymbol{v}(t_0,t)=\boldsymbol{0}$, one obtains,
\begin{align}\label{v}
\boldsymbol{v}(\tau,t)=\sum_{\alpha}\int^{\tau}_{t_0}d\tau_1\int^t_{t_0}d\tau_2\boldsymbol{u}(\tau,\tau_1)\widetilde{\boldsymbol{g}}_{\alpha}(\tau_1,\tau_2)\boldsymbol{u}^\dag(t,\tau_2).
\end{align}
In fact, $\boldsymbol{u}(\tau,t_0)$ and $\boldsymbol{v}(\tau,t)$ are related to the retarded Green function and the lesser Green function
of the device system in the nonequilibrium Green function techniques \cite{Haug2008,Wingreen1993,Jauho1994}, and Eq.~(\ref{v})
gives the generalized nonequilibrium fluctuation-dissipation theorem in the time-domain \cite{Zhang2012}.

Also, one can obtain the reduced density matrix $\rho(t)$ by solving the exact master equation (\ref{master eq}).
Denoting the empty state with $|0\rangle$, the state that one electron occupied in dot $i$ with $\left|i\right>$ ($i=1,2$),
and the doubly occupied state by $|d\rangle$,
the elements of reduced density matrix at the later time $t$ for an arbitrary initial DQD state, solving from the master equation (\ref{master eq}), can be expressed as follows \cite{Liu2016},
\begin{subequations}
\label{rho_element}
\begin{align}
&\rho_{00}(t) = A(t)\Big\{ \rho_{00}(t_{0})+\rho_{dd}(t_{0}){\rm det}\left[\boldsymbol{J}_3(t)\right]
\notag\\&~~~~~~~~~~- \sum_{i,j}\rho_{ij}(t_{0})
\boldsymbol{J}_{3ji} (t) \Big\}, \\
&\rho_{11}(t) = 1\!-\!\rho_{00}(t)\!-\!\rho^{(1)}_{22}(t),~\rho_{12}(t) = \rho^{(1)}_{12}(t),\\
& \rho_{22}(t) = 1\!-\!\rho_{00}(t)\!-\!\rho^{(1)}_{11}(t),~ \rho_{21}(t) = \rho_{12}^{*}(t),\\
& \rho_{dd}(t) = 1-\rho_{00}(t)-\rho_{11}(t)-\rho_{22}(t),
\end{align}
\end{subequations}
while the other off-diagonal density matrix elements between the different states are all zero.
The coefficients in Eq.~(\ref{rho_element}) are $A(t)=\det[\boldsymbol{1}-\boldsymbol{v}(t,t)]$,
$\bm{J}_3(t)=\boldsymbol{u}^\dag(t,t_0)[\boldsymbol{1}-\boldsymbol{v}(t,t)]^{-1}\boldsymbol{u}(t,t_0)-\boldsymbol{1}$, and $\rho(t_0)$ is the initial reduced density matrix.
The single-particle correlation function $\rho^{(1)}_{ij}(\tau,t)$ is given by \cite{Tu2008,Jin2010,Yang2017},
\begin{align}\label{rho1}
\rho_{ij}^{(1)}(\tau,t)= [\boldsymbol{u}(\tau,t_0)\boldsymbol{\rho}^{(1)}(t_0)\boldsymbol{u}^\dag(t,t_0)+\boldsymbol{v}(\tau,t)]_{ij}.
\end{align}

On the other hand, the transient transport current of electrons flowing from lead $\alpha$ into the DQDs is defined as
\begin{align}
I_{\alpha}(t)=-e\frac{d}{dt}{\rm Tr}_{S\otimes E}[~\rho_{tot}(t)N_{\alpha}],
\end{align}
where $N_{\alpha}=\sum_k c^\dag_{\alpha k}c_{\alpha k}$ is the total electron number in lead $\alpha$. Using the master equation approach,
the transport current can also be expressed in terms of the same Green functions of the DQDs, $\boldsymbol{u}(\tau,t_0)$ and $\boldsymbol{v}(\tau,t)$, as follows \cite{Tu2008,Jin2010,Yang2017} :
\begin{align}
\label{current}
I_{\alpha}(t) \! = -\frac{2e}{\hbar}{\rm Re}\int^t_{t_0}{\rm Tr}\big[\boldsymbol{g}_{\alpha}(t,\tau)\boldsymbol{\rho}^{(1)}(\tau,t)-
\! \widetilde{\boldsymbol{g}}_{\alpha}(t,\tau)\boldsymbol{u}^\dag(t,\tau) \big].
\end{align}
Indeed, the transient transport current (\ref{current}) obtained from the master equation has exactly the same
form as that derived from Keldysh's nonequilibrium Green function
technique \cite{Haug2008,Wingreen1993,Jauho1994}, except for the dependence of initial conditions,
i.e. the first term in Eq.~(\ref{rho1}),
which was omitted in Ref.~\cite{Haug2008,Wingreen1993,Jauho1994}.

Through the reduced density matrix and the transient transport current, we now can study the transient quantum
transport in the double dot AB interferometer with back-action effects from the contacts, where non-Markovian dissipation
and decoherence are fully taken into account.

\subsection{Transient linear conductance and reduced density matrix in the molecular basis of the DQDs}
Without loss of generality, we set the energy of each dot $\varepsilon_{11}=\varepsilon_{22}=\varepsilon_0$,
and the tunneling matrix between the two dots $\varepsilon_{12}=\varepsilon_{21}=-t_c$.
In order to study the molecular states of the DQDs, we may change the basis by diagonalizing $H_{DQD}$.
Denoting the bonding state (BS) and the antibonding state (AS) with the signs $-$ and $+$ respectively, the Hamiltonian of the DQDs becomes,
\begin{align}
H_{DQD}=\sum_{\nu=\pm}\varepsilon_{\nu}d_{\nu}^{\dagger}d_{\nu},
\end{align}
where $\varepsilon_{\pm}=\varepsilon_0\pm t_c$ are the corresponding energy levels,
and $d_{\pm}$ ($d_{\pm}^{\dagger}$) are the associated annihilation (creation) operators, which are given by:
\begin{align}
\begin{pmatrix} d_+ \\ d_- \end{pmatrix} = \frac{1}{\sqrt{2}}\begin{pmatrix} 1 & -1 \\ 1 & 1\end{pmatrix}\begin{pmatrix} d_1 \\ d_2 \end{pmatrix}.
\end{align}
The tunneling Hamiltonian between the molecular states and the leads is transformed to,
\begin{align}
H_T = \sum_{\alpha=L,R}\sum_{\nu=\pm}\sum_{k}(V_{\nu \alpha k} d_{\nu}^{\dagger} c_{\alpha k}+ h.c.),
\end{align}
with the tunneling matrix elements,
\begin{align}
\begin{pmatrix} V_{+ \alpha k} \\ V_{- \alpha k} \end{pmatrix}
= \frac{1}{\sqrt{2}}
\begin{pmatrix} 1 & -1 \\ 1 & 1 \end{pmatrix}
\begin{pmatrix} V_{1 \alpha k} \\ V_{2 \alpha k} \end{pmatrix}.
\end{align}
In order to observe a distinct feature of Fano resonances,
one needs to prepare a constant background coupled to a discrete level,
so we take the spectral density of lead $\alpha$ is energy independent,
$\boldsymbol{\Gamma}_{\alpha}(\omega)=\boldsymbol{\Gamma}_{\alpha}$ (i.e. the wide band limit) with
the level-width of the left lead $\Gamma_{L11}=\Gamma_{L22}=\Gamma_{L}$ and the right lead
$\Gamma_{R11}=\Gamma_{R22}=\Gamma_{R}$. Also the indirect interdot couplings through the left lead
$\Gamma_{L12}=\Gamma_{L}e^{i\frac{\varphi}{2}}$ and the right lead $\Gamma_{R12}=\Gamma_{R}e^{-i\frac{\varphi}{2}}$.
Therefore, the level-width matrix $\boldsymbol{\Gamma}_{\alpha}$ in the molecular basis is given by
\begin{align}
\begin{pmatrix}\Gamma_{++} & \Gamma_{+-} \\ \Gamma_{-+} & \Gamma_{--} \end{pmatrix}_{L,R}
=\Gamma_{L,R}\left(\boldsymbol{1}-\vec{\alpha}_{L,R}\cdot\vec{\boldsymbol{\sigma}}\right),
\end{align}
where $\vec{\alpha}_{L,R}\!=\!(\alpha_{L,R}^{x},\alpha_{L,R}^{y},\alpha_{L,R}^{z})\!=\!
(0,\pm\sin\frac{\varphi}{2},\cos\frac{\varphi}{2})$ and $\vec{\boldsymbol{\sigma}}$ are the Pauli matrices.

The transient linear conductance is defined as,
\begin{align}
\label{tranG_def}
\mathcal{G}(t)=\frac{\partial}{\partial V}I(t)|_{V=0},
\end{align}
where $I(t)=1/2~[I_L(t)-I_R(t)]$ is the net transport current and $\mu_{L,R}=E_F\pm eV/2$, in which $E_F$ denotes the Fermi level and $V$ is the bias applied to the two leads.
From Eq.~(\ref{current}) and (\ref{tranG_def}), the transient linear conductance in wide band limit is obtained as,
\begin{widetext}
\begin{align}\label{tranG}
\mathcal{G}(t)=-\frac{e^2}{h}{\rm ReTr}&\Big\{\frac{1}{4}\big(\boldsymbol{\Gamma}_L-\boldsymbol{\Gamma}_R\big)\int^t_{t_0}d\tau_1\int^t_{t_0}d\tau_2\boldsymbol{u}(t,\tau_1)
\big(\boldsymbol{\Gamma}_L-\boldsymbol{\Gamma}_R\big)\int d\omega\frac{\beta/2}{1+\cosh[\beta(\omega-E_F)]}e^{-i\omega(\tau_1-\tau_2)}\boldsymbol{u}^\dag(t,\tau_2)\notag\\
&-\frac{1}{2}\big(\boldsymbol{\Gamma}_L+\boldsymbol{\Gamma}_R\big)\int^t_{t_0}d\tau\int d\omega\frac{\beta/2}{1+\cosh[\beta(\omega-E_F)]}e^{-i\omega(t-\tau)}\boldsymbol{u}^\dag(t,\tau)\Big\},
\end{align}
\end{widetext}
where the retarded Green function $\boldsymbol{u}(t,\tau)$ has the following solution,
\begin{align}\label{u}
\boldsymbol{u}(t,\tau)  =
 \exp\!\Big[-\Big(i\boldsymbol{\varepsilon}+\frac{1}{2}\boldsymbol{\Gamma} \Big)(t-\tau)\Big]\!,
\end{align}
and $\boldsymbol{\varepsilon}=\begin{pmatrix} \varepsilon_{+}& \!\! 0\\0& \varepsilon_{-}\end{pmatrix}$ and $\boldsymbol{\Gamma}=\boldsymbol{\Gamma}_{L}+\boldsymbol{\Gamma}_{R}$.
When the leads are initially at zero temperature ($\beta\rightarrow\infty$),
the frequency dependent term in Eq.~(\ref{tranG}) is reduced to a delta function:
$\frac{\beta/2}{1+\cosh[\beta(\omega-E_F)]}\rightarrow \delta(\omega-E_F)$.
Then in the steady-state limit ($t\rightarrow\infty$), one can reproduce the linear conductance obtained previously in Ref. \cite{Kubala2002,Ladron2003,Kang2004,Bai2004,Lu2005,Kubo2006},
\begin{align}
\label{steadyG}
\mathcal{G}=\frac{e^2}{h}\textrm{Tr}\left[\boldsymbol{\Gamma}_{L}\boldsymbol{G}^{R}(E_F)\boldsymbol{\Gamma}_{R}\boldsymbol{G}^{A}(E_F)\right],
\end{align}
where the retarded Green function in energy domain,
\begin{align}
\mathbf{G}^{R}(E_F) & = -i\lim_{t\rightarrow\infty}\lim_{t_0\rightarrow-\infty}\int_{t_0}^{t}e^{iE_F (t-\tau)}\boldsymbol{u}(t,\tau)d\tau, \notag \\
 & = \Big(  E_F \boldsymbol{I}-  \boldsymbol{\varepsilon}+ \frac{i}{2} \boldsymbol{\Gamma} \Big)^{-1},
\end{align}
and the advanced Green function $\mathbf{G}^{A}(E_F)=[\mathbf{G}^{R}(E_F)]^\dag$.

In the molecular basis,
the elements of reduced density matrix for an arbitrary initial DQD state are expressed as \cite{Liu2016},
\begin{subequations}
\begin{align}
&\rho_{00}(t) = A(t)\Big\{ \rho_{00}(t_{0})+\rho_{dd}(t_{0}){\rm det}\left[\boldsymbol{J}_3(t)\right]
\notag\\&~~~~~~~~~~- \!\!\!\!\!\sum_{\nu,\nu'= \pm}\!\!\! \rho_{\nu\nu'}(t_{0})
\boldsymbol{J}_{3\nu'\nu} (t) \Big\}, \\
&\rho_{++}(t) = 1\!-\!\rho_{00}(t)\!-\!\rho^{(1)}_{--}(t),~\rho_{+-}(t) = \rho^{(1)}_{+-}(t),\\
& \rho_{--}(t) = 1\!-\!\rho_{00}(t)\!-\!\rho^{(1)}_{++}(t),~ \rho_{-+}(t) = \rho_{+-}^{*}(t),\\
& \rho_{dd}(t) = 1-\rho_{00}(t)-\rho_{++}(t)-\rho_{--}(t).
\end{align}
\end{subequations}
If the dots are initially in an empty state, the elements of the reduced density matrix are simplified to,
\begin{subequations}
\label{rho_em}
\begin{align}
&\rho_{00}(t)= \det[\boldsymbol{1}-\boldsymbol{v}(t,t)], \\
&\rho_{++}(t) = 1\!-\!\rho_{00}(t)\!-\!\boldsymbol{v}_{--}(t,t),~~\rho_{+-} = \boldsymbol{v}_{+-}(t,t),\\
& \rho_{--}(t) = 1\!-\!\rho_{00}(t)\!-\!\boldsymbol{v}_{++}(t,t),~~\rho_{-+} = \boldsymbol{v}_{+-}^{*}(t,t),\\
& \rho_{dd} = \det[\boldsymbol{v}(t,t)],
\end{align}
\end{subequations}
Thus, the reduced density matrix elements of the DQDs are also fully determined by the Green function solution Eq.~(\ref{v})
through the solution Eq.~(\ref{u}).

\section{Buildup of Fano resonance in the time domain}

\subsection{Steady-state Fano resonance}
To get an instructive physical picture, we shall present first the steady-state result of the Fano resonance in this
double-dot AB interferometer. With loss of generality, we focus on the case of the symmetric geometry $\Gamma_L=\Gamma_R=\Gamma/2$.
Then, according Eq.~(\ref{steadyG}), the steady-state linear conductance at zero temperature is explicitly given by,
\begin{align}\label{G_Kang}
\mathcal{G}=\frac{e^2}{h}\frac{(e_+-e_-)^2}{(e^2_++1)(e^2_-+1)}.
\end{align}
The dimensionless energies $e_{+,-}\equiv(E_F-\varepsilon_{+,-})/(\Gamma_{+,-}/2)$, with
$\Gamma_+=\Gamma_{L++}+\Gamma_{R++}=2\Gamma\sin^2(\varphi/4)$ and
$\Gamma_-=\Gamma_{L--}+\Gamma_{R--}=2\Gamma\cos^2(\varphi/4)$ which are sensitive to the AB phase.
Equation~(\ref{G_Kang}) reproduces the results in Ref.~\cite{Kang2004},
and reveals two resonances,
a Breit-Wigner resonance and a Fano resonance,
when the bandwidth
of the two molecular states $\Gamma_+\gg\Gamma_-$ or $\Gamma_-\gg\Gamma_+$.
Explicitly, when $\Gamma_+\gg\Gamma_-$, in the energy scale near the antibonding energy ($|e_-|\gg 1$),
the conductance of Eq.~(\ref{steadyG}) follows the Breit-Wigner form of its width $\Gamma_+/2$ \cite{Kang2004}:
\begin{align}
\mathcal{G}\simeq\frac{e^2}{h} \frac{1}{e^2_++1}.
\end{align}
While in the energy scale near the bonding energy ($|e_-|\lesssim 1$), one can find that the conductance shows Fano resonance profile:
\begin{align}
\mathcal{G}\simeq \frac{e^2}{h}\frac{1}{q^2+1}\frac{(e_-+q)^2}{e^2_-+1},
\end{align}
where the Fano factor $q=4t_c/\Gamma_+$ which is the ratio of the interdot coupling and the level broadening of the antiboding state.
Thus, the conductance is composed of a Breit-Wigner
resonance at the antibonding energy and a Fano resonance near the bonding energy.
If one takes $\Gamma_-\gg \Gamma_+$, the same analysis is applied with the
role of the bonding and antibonding states interchanged with the Fano factor given by $q=-4t_c/\Gamma_-$.

\begin{figure}
\centerline{\scalebox{0.27}{\includegraphics{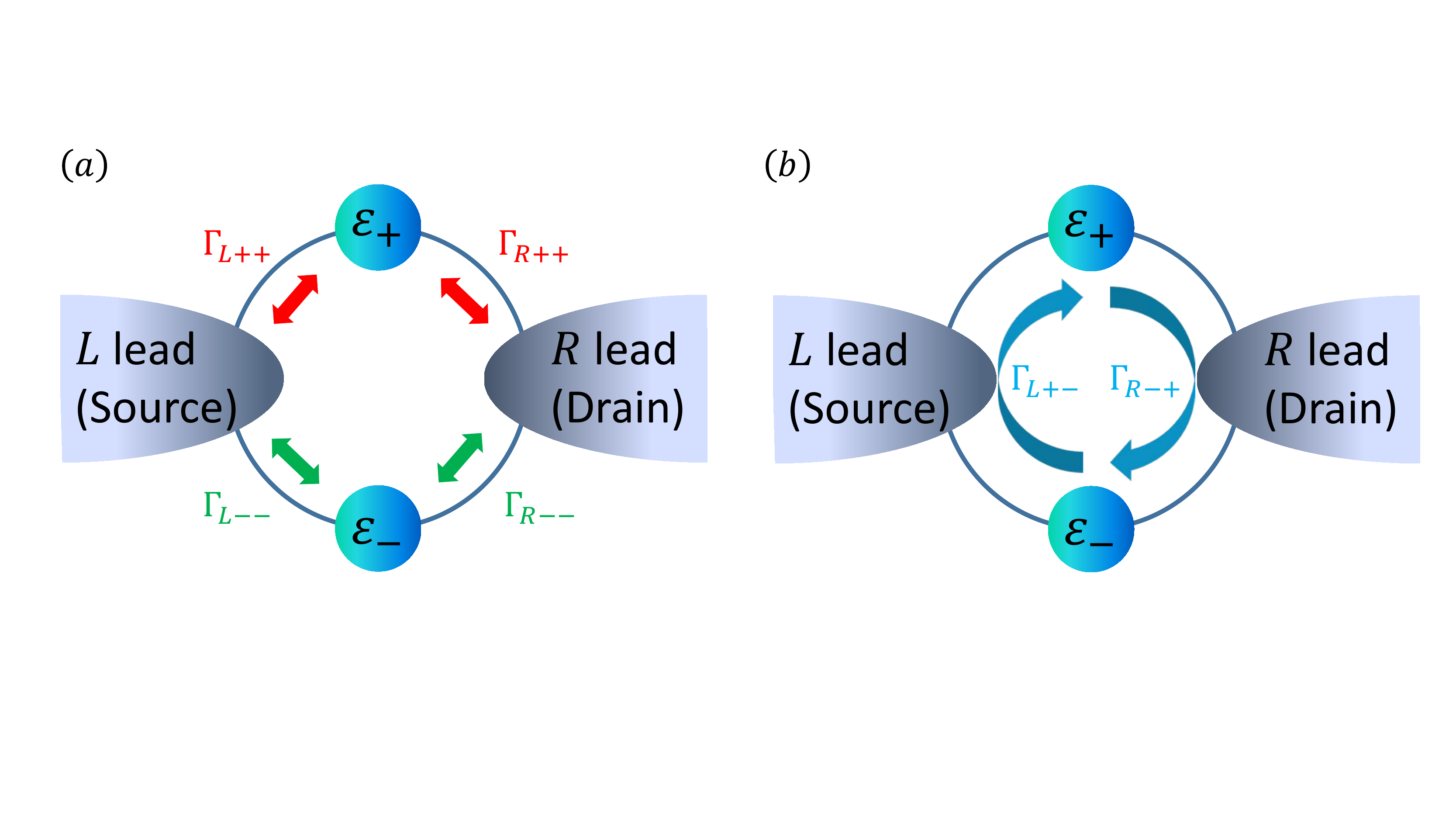}}}
\caption{Different paths for electron transport between the source and drain in the symmetric situation.
(a)Electrons counter transport through bonding and antibonding channels, respectively.
(b)Electrons transport through both bonding and antibonding channels.}
\label{G_diff_path}
\end{figure}

To get a better understanding of the electron transport behavior through different paths,
we divide the linear conductance Eq.~(\ref{steadyG}) into components flowing through
the bonding and antibonding state channels,
plus the interference of electron transport through different paths in the AB ring geometry,
\begin{align}
\mathcal{G}=\mathcal{G}_++\mathcal{G}_-+\mathcal{G}_{+-}.
\end{align}
In this case, the components are reduced to
\begin{subequations}\label{steadyG'}
\begin{align}
\mathcal{G}_{\pm}=&\frac{e^2}{h}\Gamma_{L\pm\pm}\Gamma_{R\pm\pm}\left|G_{\pm\pm}^{R}(E_F)\right|^{2}=\frac{e^2}{h}\frac{1}{e^2_\pm+1}, \\
\mathcal{G}_{+-}=& ~2\frac{e^2}{h}\mathrm{Re}\big\{\Gamma_{L+-}G_{--}^{R}\Gamma_{R-+}G_{++}^{A}\big\}\notag\\
=&-\frac{e^2}{h}\frac{2(e_+e_-+1)}{(e^2_++1)(e^2_-+1)},
\end{align}
\end{subequations}
where $\mathcal{G}_{\pm}$ are respectively the conductance of electron counter
flowing through the bonding and antibonding state channels as shown in Fig.~\ref{G_diff_path}(a),
which gives two Breit-Wigner resonances in the electron conductance. On the other hand,
because there is no direct coupling between the bonding and antibonding states
(Green function $\boldsymbol{u}(t,t_0)$ is diagonal),
$\mathcal{G}_{+-}$ describes interference between electron transport through bonding
and antibonding channels, induced by the indirect
interdot coupling through the two leads as shown in Fig.~\ref{G_diff_path}(b). This interference leads to Fano resonance.
Thus, it is obvious that when there is no indirect interdot coupling, Fano resonance will be suppressed, which results in
two Breit-Winger peaks in the steady-state conductance as shown in Ref.~\cite{Kubo2006}.

\begin{figure}
\centerline{\scalebox{0.4}{\includegraphics{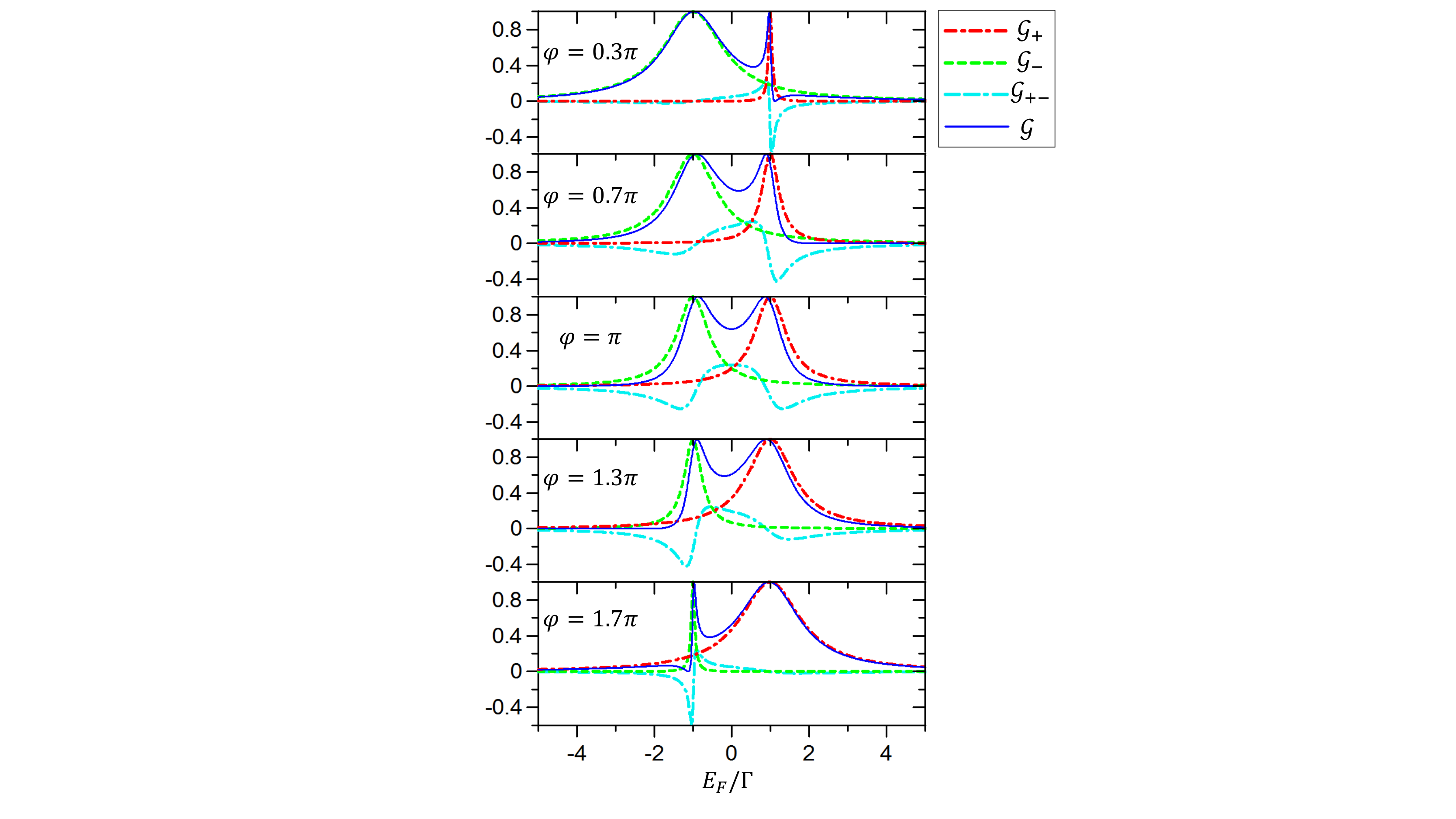}}}
\caption{Components of steady linear conductance at zero temperature as a function of Fermi energy for different $\varphi$, where $\varepsilon_0=0$, $t_c=1/\Gamma$.}
\label{G_steady}
\end{figure}

Figure~\ref{G_steady} shows the conductances as a function of Fermi energy of the leads for different values
of $\varphi$ which solely determines the band width and the Fano factor when one fixes the interdot coupling $t_c$.
One can see that conductance terms $\mathcal{G}_+$ and $\mathcal{G}_-$ contribute two Breit-Winger peaks at the bonding and antibonding energies as discussed above.
Meanwhile, the interference term $\mathcal{G}_{+-}$ is every sensitive to the AB phase.
For $\varphi=0.3\pi$, $\Gamma_-\gg\Gamma_+$,
the interference term makes a destructive and a constructive contributions to the resonance transport to the antibonding state,
which induces a Fano resonance at the antibonding state energy in the total condutance.
The interference pattern shows full constructive and destructive interferences around resonance frequency $\varepsilon_+$.
As a result, the total conductance of the system shows a Breit-Wigner resonance
and a Fano resonance at the bonding and antibonding state energies, respectively.
When the AB phase changes, for example, $\varphi=0.7\pi$, the interference term $\mathcal{G}_{+-}$ still leads to Fano resonance
at the antibonding state except that factor $q$ gets larger because of the smaller $\Gamma_-$.
When $\varphi=\pi$, $\Gamma_-=\Gamma_+$, the interference term $\mathcal{G}_{+-}$ is symmetric with the Fermi energy $E_F$.
Actually, it induces two Fano resonances around the bonding and antiboding state energies with $q=-4$ and $q=4$ in the total conductance, respectively.
On the other limit, $\varphi=1.3\pi$ and $\varphi=1.7\pi$ for $\Gamma_+\gg\Gamma_-$, the role of the bonding
and antibonding states interchanges, such that the conductance consists of
a Breit-Wigner resonance at the antibonding state energy, and a Fano resonace at the bonding state energy.

\subsection{Buildup of Fano resonance in the time-domain}
\begin{figure}
\centerline{\scalebox{0.35}{\includegraphics{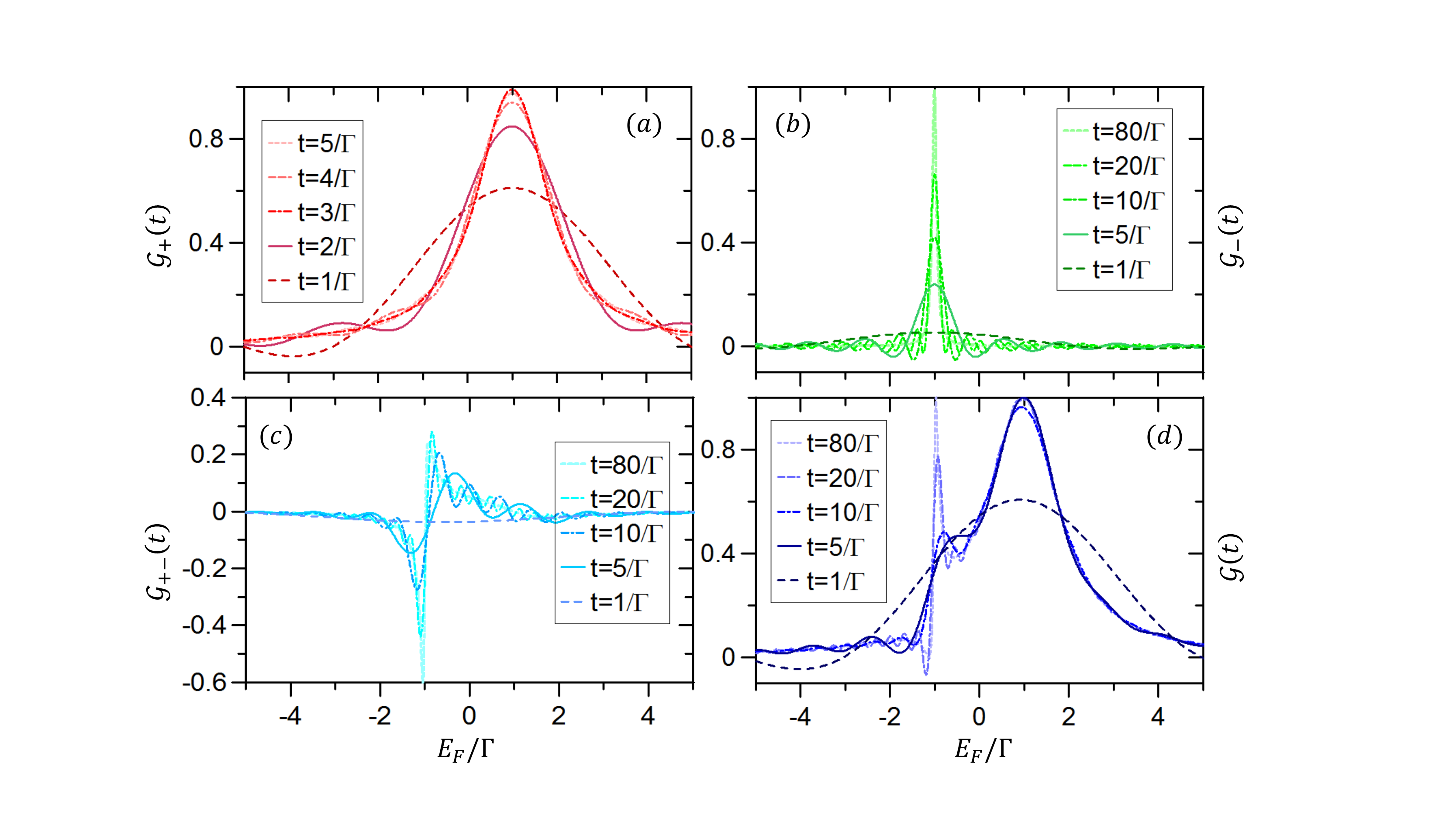}}}
\caption{Time evolution of the components of steady linear conductance at zero temperature as a function of Fermi energy, where $\varepsilon_0=0$, $t_c=1/\Gamma$ and $\varphi=1.7\pi$.}
\label{G_kT=0}
\end{figure}

After examining the above general steady-state resonance behaviors, we now explore the real-time dynamics of the buildup of Fano resonance. We set $t_0=0$ hereafter.
The time evolution of the transient linear conductance at zero temperature with the AB phase $\varphi=1.7\pi$ ($\Gamma_+\gg\Gamma_-$) is displayed in Fig.~\ref{G_kT=0}.
At the beginning ($t=1/\Gamma$), because the coupling of the antibonding state to the two leads is much larger than that of the bonding state,
electrons mainly transport to the antibonding state, although its energy is higher than
that of the bonding state. As one can see from Fig.~\ref{G_kT=0}, when $t=1/\Gamma$, the component $\mathcal{G}_+$
grows faster and forms a resonance at the antibonding state energy, while $\mathcal{G}_-$ and
$\mathcal{G}_{+-}$ are almost zero.
As time increasing, $\mathcal{G}_+$ (Fig.~\ref{G_kT=0}(a)) rapidly steadies to a Breit-Winger resonance
with full transmission at the antibonding state energy within $t=5/\Gamma$.
The time scale to reach the steady state for the Breit-Wigner resonance is determined by the Green function
component $u_{++}(t,t_0)$. Thus, this time scale is proportional to the inverse of the decay rate $\Gamma_+/2$ of $u_{++}(t,t_0)$ (See Eq.~(\ref{u})).
On the other hand, the steady state of $\mathcal{G}_-$ and $\mathcal{G}_{+-}$ is reached in the scale of the inverse of the decay rate $\Gamma_-/2$.
The formation of Fano resonance at the bonding state energy comes from the combination of these two terms in this case.
Therefore, the time scale of the formation of the Fano resonance is determined by the coupling
between the bonding state and the two leads, which is much longer than the time scale of the
formation of the Breit-Wigner resonance. In Fig.~\ref{G_kT=0}(d), the time evolution of the total conductance is presented, and as one can see the
Fano resonance is fully buildup around $t \simeq 80/\Gamma$, where
the interference pattern combines both the full constructive and destructive interferences.

\begin{figure}
\centerline{\scalebox{0.35}{\includegraphics{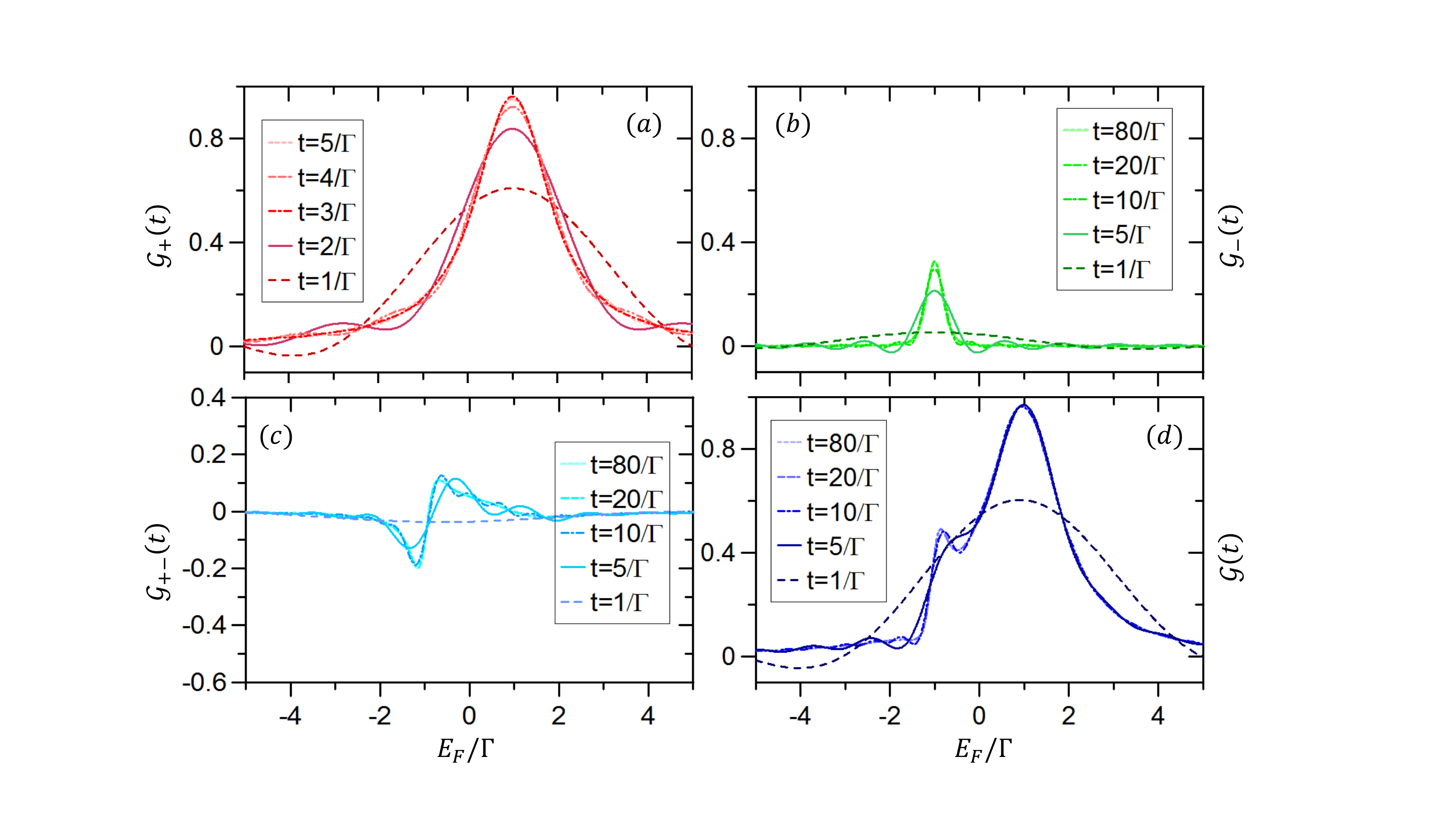}}}
\caption{Time evolution of the components of steady linear conductance as a function of Fermi energy at initial temperature of lead $k_BT=0.1\Gamma$, where $\varepsilon_0=0$, $t_c=1/\Gamma$ and $\varphi=1.7\pi$.}
\label{G_kT=01}
\end{figure}

Considering the thermal effect, the transient dynamics of the linear conductance at
initial temperature of the leads $k_BT=0.1\Gamma$ is shown in Fig.~\ref{G_kT=01}.
For the conductance through the antibonding channel (Fig.~\ref{G_kT=01}(a)),
the result is almost the same as that in the zero temperature case, except
that the transmission at the antibonding energy is slightly decreased due to the thermal effect.
In this case, the level width of the antibonding state $\Gamma_+/2 \backsimeq 0.95\Gamma$ (for $\varphi=1.7\pi$) still dominates the damping, while
the temperature induced level broadening is proportional to $k_BT=0.1\Gamma$.
Hence, the resonant transport through the antibonding channel is less affected by thermal broadening effect.
For the conductance passing bonding channels and the interference between the bonding and antibonding channels (Fig.~\ref{G_kT=01}(b) and \ref{G_kT=01}(c)),
the resonant behavior depends mainly on level width $\Gamma_-/2\backsimeq 0.05\Gamma$ (for $\varphi=1.7\pi$)
which is less than the thermal broadening effect ($k_BT=0.1\Gamma$).
Thus, the amplitudes of the resonance peak of $\mathcal{G}_-$ and the conductance $\mathcal{G}_{+-}$ are significantly suppressed by the thermal effect.
Moreover, the time scale to the steady state of conductance $\mathcal{G}_-$ and $\mathcal{G}_{+-}$ is also largely reduced due to the thermal fluctuations.
As a result, the Fano resonance is strongly suppressed as one can see from the total linear conductance $\mathcal{G}$ in Fig.~\ref{G_kT=01}(d),
where the full constructive and destructive inferences in the
Fano resonance are significantly reduced.

\subsection{The electron distributions in the molecular states associated with Fano resonance in terms of the reduced density matrix}

\begin{figure*}
\hspace*{-15mm}
\vspace*{-3cm}
\centerline{\scalebox{0.55}{\includegraphics{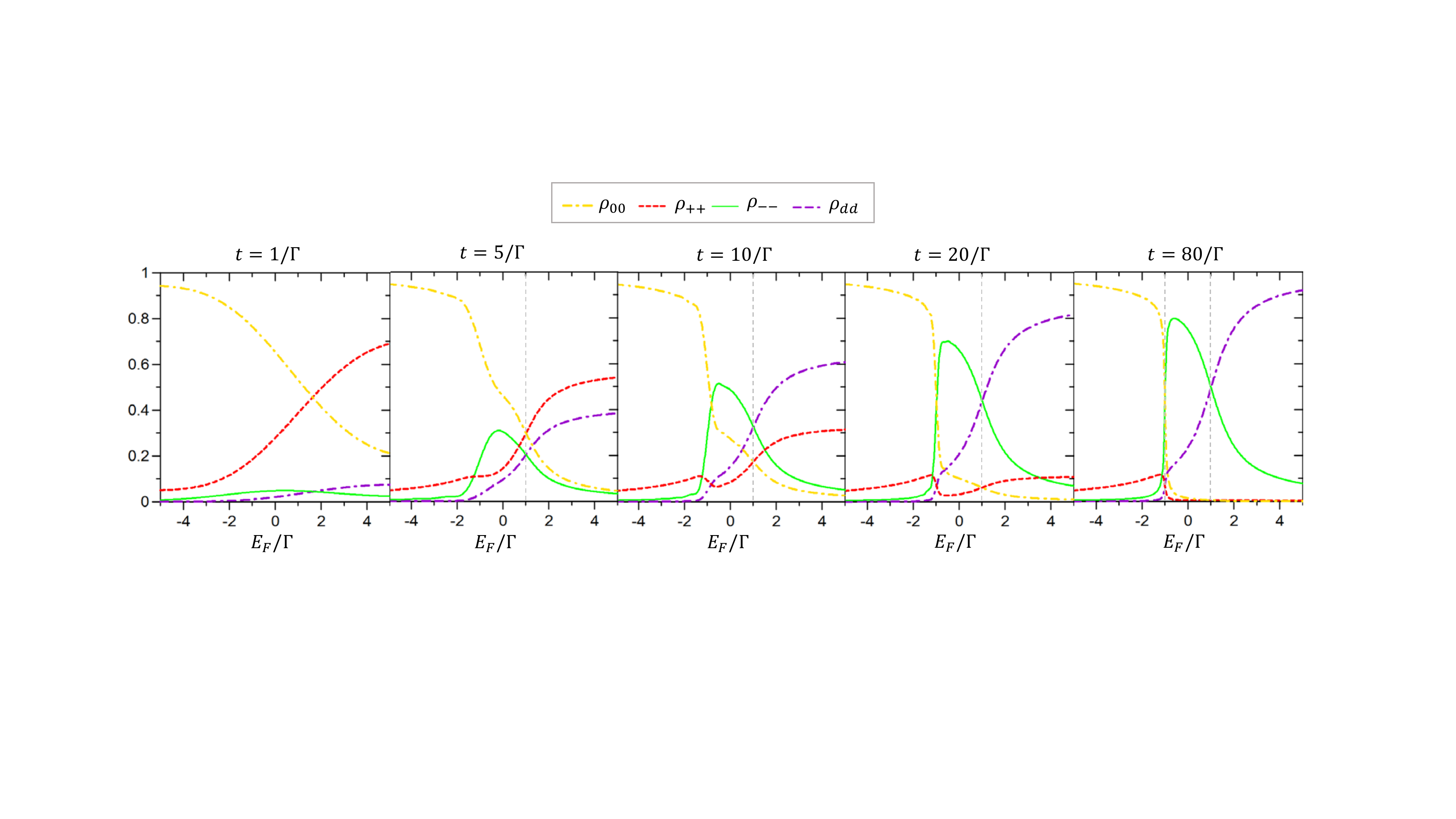}}}
\caption{Time evolution of the elements of the reduced density matrix of the DQDs at zero temperature,
where $\varepsilon_0=0$, $t_c=1/\Gamma$ and $\varphi=1.7\pi$.
The vertical dashed lines denote the happening of resonances.}
\label{rho_kT=0}
\end{figure*}

To see how the state of the DQD system changes when resonances happen,
we turn to explore the dynamics of the reduced density matrix of the DQDs.
Consider the system initially in an empty state,
the elements of the reduced density matrix are given by Eq.~(\ref{rho_em}),
which are fully determined by Green function $\boldsymbol{v}(t,t)$.
In the symmetric setup of the DQDs ($\Gamma_L=\Gamma_R=\Gamma/2$),
the off-diagonal terms of the reduced density matrix in the molecular basis
are zero because Green function $\boldsymbol{v}(t,t)$ is diagonal.
Moreover, we let the leads be unbiased, i.e. $\mu_L=\mu_R=E_F$,
the transient dynamics of the reduced density matrix at
zero temperature with AB phase $\varphi=1.7\pi$ ($\Gamma_+\gg\Gamma_-$) is shown in Fig.~\ref{rho_kT=0}.
In the beginning ($t=1/\Gamma$), electron transport is dominated by the coupling between the DQDs and the leads.
As $E_F$ varying from negative to positive energy by gate control of the dot energy $\varepsilon_0$, electrons prefer to occupy the antibonding state because of its
larger coupling to the leads, so that the empty state probability $\rho_{00}$ is
decreased and the antibonding state probability $\rho_{++}$ is increased while $\rho_{--}$ and $\rho_{dd}$ almost keep to zero.
With time going ($t\geqslant 5/\Gamma$), electrons transport from the leads to the DQDs
will also be trapped in the bonding state when $E_F>\varepsilon_-$, which results in the increase of the probability of the doubly occupied state as well.
As one can see, later than $t=5/\Gamma$, $\rho_{--}$ dramatically increases after $E_F$
passing the bonding state energy $\varepsilon_-$ and then decreases after reaching a maximum value.
The maximum value gets larger with time going as shown in Fig.~\ref{rho_kT=0}.
The probability of the doubly occupied state $\rho_{dd}$ also increases when $E_F>\varepsilon_-$ and keeping climbing with $E_F$.
While the probability of the antibonding state is decreased with time after $E_F$ passing bonding state energy $\varepsilon_-$.

\begin{figure*}
\hspace*{-15mm}
\vspace*{-3cm}
\centerline{\scalebox{0.55}{\includegraphics{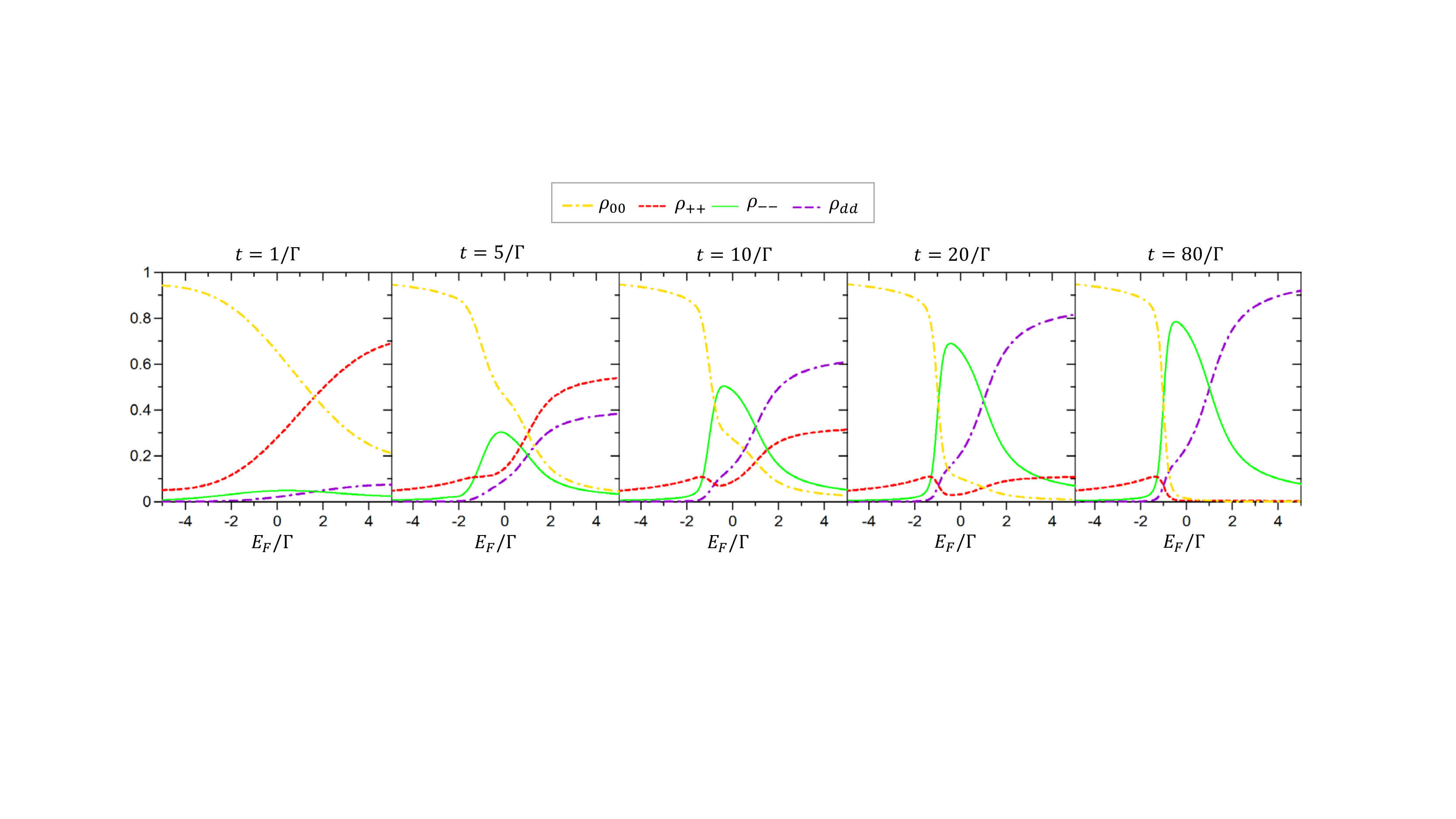}}}
\caption{Time evolution of the elements of the reduced density matrix of the DQDs at zero temperature, where $\varepsilon_0=0$, $t_c=1/\Gamma$ and $\varphi=1.7\pi$.}
\label{rho_kT=01}
\end{figure*}

Eventually, when the total system reaches the steady-state limit ($t=80/\Gamma$),
we can see that in the range $E_F<\varepsilon_-$, the state of the DQD system has the most probability in an empty state ($\rho_{00} \backsimeq 1$),
while there are a few probabilities for electrons to occupy in the antibonding state because of its larger coupling to the leads.
At $E_F=\varepsilon_-$ ($\varepsilon_-=-\Gamma$ in Fig.~\ref{rho_kT=0}), there are two abrupt transitions for the occupation probabilities. One is between the empty
state $\rho_{00}$ and the bonding state $\rho_{--}$, the other one is between the antibonding state $\rho_{++}$ and the doubly occupied state $\rho_{dd}$.
Both of the transitions imply that there is a resonant tunneling from the leads to the bonding state.
In the range $E_F>\varepsilon_-$, the probabilities $\rho_{00}$ and $\rho_{++}$ rapidly reduce to zero.
As mentioned, the probability $\rho_{--}$ reaches a maximum value and then decreases in large $E_F$,
while $\rho_{dd}$ keeps climbing with $E_F$,
which leads to a crossover between
the bonding state and doubly occupied state at $E_F=\varepsilon_+$  ($\varepsilon_+=\Gamma$ in Fig.~\ref{rho_kT=0}).
Compared with the transient conductance in Fig.~\ref{G_kT=0}, we find that resonances happens when there is a transition of the occupation probabilities between
the states differing with one electron, e.g. $\left|0\right>\leftrightarrow\left|\pm\right>$ or $\left|\pm\right>\leftrightarrow\left|d\right>$.
The transition with gentle slope (namely a crossover) leads to a Breit-Wigner resonance, whereas the abrupt transition leads to a Fano resonance.
In particular, in Fig.~\ref{rho_kT=0}, the Breit-Wigner resonance at the antibonding state energy
happens when crossovers happen from $\rho_{00}$ to $\rho_{++}$ and from $\rho_{--}$ to $\rho_{dd}$.
Fano resonance happens
at the bonding state energy when probability transits abruptly from $\rho_{00}$ to $\rho_{--}$ and from $\rho_{++}$ to $\rho_{dd}$.
%If one takes $\Gamma_-\gg\Gamma_+$ (e.g. $\varphi=0.3\pi$), one should observe a cross over between
%$\rho_{00}$ and $\rho_{--}$ at the bonding energy and two abrupt transition between
%the bonding state and doubly occupied state, empty state and the antibonding state at the antibonding energy.
However, when the thermal effect is considered,
the elements of reduced density matrix turn into the ones shown in Fig.~\ref{rho_kT=01}.
Figure ~\ref{rho_kT=01} shows that the abrupt transitions at the bonding state energy is apparently eased off because of thermal effects.
In other words, the thermal effects smear Fano resonance.

\section{Summary}
In summary, using the quantum transport theory based on the master equation approach \cite{Tu2008,Jin2010,Yang2017},
we obtain the transient linear conductance and the elements of reduced density matrix of
the DQDs not only at zero temperature but also at finite temperature.
The conductance can be divided into components flowing through
the bonding and antibonding state channels, plus the in-
interference of electron transport through different paths
in the AB ring geometry.
We investigate the transient dynamics in terms of each component of linear conductance.
We find that the time scale for the formation of Fano resonance is much slower than
the formation of the Breit-Wigner resonance because the large differences of the coupling between
the each level of DQDs and the leads. Therefore, Fano resonance can be smeared easily by the thermal effect.
From the transient dynamics of the reduced density matrix,
we find that resonances happens when there is a transition of the occupation probabilities between
the states differing by one electron, e.g. $\left|0\right>\leftrightarrow\left|\pm\right>$ or $\left|\pm\right>\leftrightarrow\left|d\right>$.
The transition with gentle slope (corresponding to a crossover) leads to a Breit-Wigner resonance, whereas the abrupt transition leads to a Fano resonance.
This abrupt transition will be eased up with thermal effect comes in.
The real-time dynamics of electron conductance in the double-dot AB interferometer help us to understand origin of the formation of
Fano resonance which are expected to be observed in future experiments.

\section{Acknowledgements}
We thank Amnon Aharony and Ora Entin-Wohlman for helpful discussions and suggestions on the Manuscript.
This work is supported by the Ministry of Science and Technology of the Republic of China under the contract No. MOST 105-2112-M-006-008-MY3

\end{document}